\documentclass[floats,floatfix,showpacs,amssymb,prd,twocolumn,superscriptaddress,nofootinbib]{revtex4-1}

\usepackage{amssymb,amsmath,verbatim,mathtools,needspace,enumitem,etoolbox,graphicx,physics,microtype,afterpage,bm}

\usepackage[dvipsnames, usenames]{xcolor}

\definecolor{linkcolor}{rgb}{0.0,0.3,0.5}
\usepackage[unicode, colorlinks=true, linkcolor=linkcolor, citecolor=linkcolor, filecolor=linkcolor, urlcolor=linkcolor, linktocpage, breaklinks]{hyperref}
\usepackage[all]{hypcap}
\usepackage[T1]{fontenc}
\usepackage[utf8]{inputenc}
\usepackage{tabularx}

\allowdisplaybreaks
\newcommand\prlsec[1]{\vspace{2mm}\noindent {\bf \emph{#1}}}

\def\be{\begin{equation}}
\def\ee{\end{equation}}
\newcommand{\beq}{\begin{eqnarray}}
\newcommand{\eeq}{\end{eqnarray}}

\begin{document}

\title{Hierarchy of Angular Instabilities in Scalarized Black Holes}

\author{Jose Luis Bl\'azquez-Salcedo} 
\email{jlblaz01@ucm.es}
\author{Luis Manuel Gonz\'alez-Romero} 
\email{mgromero@ucm.es}
\author{{Fech Scen Khoo}}
\email{fkhoo@ucm.es}
\affiliation{Departamento de F\'isica Te\'orica and IPARCOS, Facultad de Ciencias F\'isicas, Universidad Complutense de Madrid, Spain}
\author{{Jutta Kunz}}
\email{jutta.kunz@uni-oldenburg.de}
\affiliation{Institut f\"ur  Physik, Universit\"at Oldenburg, Postfach 2503, D-26111 Oldenburg, Germany}
\author{{Pablo Navarro Moreno}}
\email{panava03@ucm.es}
\affiliation{Departamento de F\'isica Te\'orica and IPARCOS, Facultad de Ciencias F\'isicas, Universidad Complutense de Madrid, Spain}

\date{\today}

\begin{abstract}
We investigate the stability of scalarized black holes in Einstein-scalar-Gauss-Bonnet-Ricci theory along their fundamental branches. 
We show that initially stable solutions first lose nonspherical stability in the eikonal regime, while lower multipoles remain stable. 
As the branch is continued, instability extends systematically toward lower multipoles, forming an ordered hierarchy of deformation instabilities extending down to the quadrupole mode, while the dipole sector remains stable. 
The instability thresholds obey a common scaling law and approach finite eikonal limits, defining the boundary of the angularly stable region. 
We demonstrate that the previously identified quadrupole and angular-Laplacian instabilities are connected by a continuous hierarchy of instability thresholds spanning the angular sectors of the theory. 
This hierarchy is distinct from radial stability, which changes only at branch turning points, and reveals a previously unexplored angular organization of instabilities in scalarized black holes.
\end{abstract}

\maketitle

\prlsec{Introduction.}
Determining the stability of black hole (BH) solutions is essential for understanding which branches of solutions can be physically realized and how transitions between different phases occur. 
This question is particularly important in extensions of general relativity containing additional degrees of freedom, where new families of compact objects and new instability channels can appear \cite{Berti:2015itd,Kobayashi:2019hrl,Cardoso:2019rvt,Doneva:2022ewd}.
More generally, an important open question is how instabilities emerge and spread across perturbative sectors when continuous branches of solutions lose stability.
In theories admitting scalarized BHs, this question is particularly relevant because solutions often appear in continuous families connected through bifurcations from general relativity \cite{Damour:1993hw,Doneva:2017bvd,Silva:2017uqg,Antoniou:2017acq,Antoniou:2017hxj,Bakopoulos:2020dfg,Doneva:2021tvn}.

Most stability analyses focus on a specific perturbative sector, such as radial perturbations or individual angular multipoles \cite{Blazquez-Salcedo:2018jnn,Silva:2018qhn,Blazquez-Salcedo:2022omw,Blazquez-Salcedo:2024rvb}. 
However, comparatively little is known about how different angular sectors are related to one another along a continuous BH branch. 
In particular, it remains unclear whether instability appears independently in different multipoles or whether there exists an underlying organization connecting the onset of instability across angular scales.

The large-$l$ sector is especially subtle in higher-curvature gravity. 
In general relativity, eikonal quasinormal modes are closely related to the properties of unstable null geodesics \cite{Cardoso:2008bp}. 
In higher-curvature theories, however, this correspondence can fail, and the eikonal sector may itself become unstable \cite{Konoplya:2017ymp,Konoplya:2017wot,Konoplya:2017lhs,Konoplya:2008ix}. 
Thus, large angular momentum modes are not merely an asymptotic diagnostic of the spectrum, but can control the onset of instability.

In this Letter, we address this question for scalarized BHs in Einstein-scalar-Gauss-Bonnet-Ricci (EsGBR) theory \cite{Antoniou:2020nax,Antoniou:2021zoy,Antoniou:2022agj}. 
We show that branches previously identified as partially stable possess a rich hierarchy of nonspherical deformation instabilities. 
Starting from an angularly stable region, instability first appears in the eikonal regime and subsequently extends toward progressively lower multipoles. 
The corresponding instability thresholds form a continuous hierarchy obeying a common scaling law and approaching finite eikonal limits, while the dipole sector remains stable throughout the domain studied. 
This hierarchy is distinct from radial stability, which changes only at the turning points of the branches.
We find that the quadrupole instability and the angular-Laplacian eikonal instability arise as the low- and high-$l$ limits of a common instability structure.

Our results reveal a previously unexplored angular organization of instabilities in scalarized black holes and suggest that the stability landscape of gravitational solution branches may possess a richer structure than indicated by analyses restricted to individual perturbative sectors.

\prlsec{EsGBR Theory.}
We briefly recall the scalarized BH solutions for the EsGBR action \cite{Antoniou:2020nax,Antoniou:2021zoy,Antoniou:2022agj}
\begin{eqnarray}  
\label{act}
 {\cal S} = \frac{1}{16 \pi}  \int  \mathrm{d}^4x \sqrt{-g}  \left[ R - \frac{1}{2}(\partial_\mu \phi)^2 \right. 
 \nonumber\\
\left.  -   
{f(\phi)}
\left(\frac{\beta}{2}R -\alpha R_{\mathrm{GB}}^2\right)  \right]
\end{eqnarray} 
with a real scalar field $\phi$, 
coupling constants $\alpha$ and $\beta$, and GB invariant
 $R^2_{\rm GB} = R_{\mu\nu\rho\sigma} R^{\mu\nu\rho\sigma}
- 4 R_{\mu\nu} R^{\mu\nu} + R^2$ \,. {The coupling function is $f(\phi)=\phi^2/2$.}
EsGBR theories are well motivated from a cosmological point of view, since they allow GR to be a cosmological attractor \cite{Antoniou:2020nax,Antoniou:2021zoy,Ventagli:2021ubn}:
no fine-tuning of the scalar field in the early Universe is needed in order to have a vanishing scalar field at late times.
Related studies have explored the dynamics of these theories in gravitational collapse, nonlinear evolutions, and scalar-hair formation \cite{Thaalba:2023fmq,Liu:2022eri,Doneva:2024ntw}.

Varying the action \eqref{act} with respect to the metric gives the generalized Einstein equations
\begin{align}
    E_{\mu\nu} = G_{\mu\nu} - \frac{1}{2}T_{\mu\nu}^{\mathrm{(eff)}} = 0 \,,
\end{align}
with the effective stress-energy tensor
\begin{align}
    T_{\mu\nu}^{\mathrm{(eff)}} = T_{\mu\nu}^{(\phi)} - 2\alpha T_{\mu\nu}^{(\mathrm{GB})} + \beta T_{\mu\nu}^{(\mathrm{R})}
\end{align}
receiving contributions from the scalar-field, Gauss-Bonnet, and Ricci-coupling terms, respectively. 
Variation with respect to the scalar field leads to the generalized Klein-Gordon equation
\begin{eqnarray}
\label{dil-eq}
\nabla^2 \phi - 
\left(\frac{\beta}{2}R -\alpha R_{\mathrm{GB}}^2\right) 
{f'(\phi)}
 =0 \,.
\end{eqnarray} 
This equation contains the effective mass term
\begin{equation}
    m^2_{\rm eff} =  \frac{\beta}{2} R - \alpha R^2_{\rm GB} \,,
\end{equation}
which can trigger spontaneous scalarization and gives rise to the appealing cosmological properties of the model.
The static, spherically symmetric spacetime solutions are obtained by adopting the  general static form of the metric
\begin{equation}\label{bg_metric}
ds^2=-e^{2\nu(r)}dt^2+e^{2\lambda(r)}dr^2+r^2 \left( d\theta^2+\sin^2{\theta}d\varphi^2 \right) \,,
\end{equation}
and solved subject to regularity, asymptotic flatness, and a vanishing scalar field at infinity.
Regularity at the origin yields a constraint between the horizon radius $r_H$, the scalar field at the horizon 
$\phi_H$
and the coupling constants.
The BH mass $M$ and the scalar charge $D$ follow from the asymptotic fall-off of the functions.

Depending on the coupling constants, the theory allows for radially stable fundamental branches of scalarized BHs, already for the simplest coupling function $f({\phi})={\phi}^2/2$ that gives rise to spontaneous scalarization \cite{Antoniou:2021zoy,Antoniou:2022agj}. 
These scalarized BHs bifurcate from the Schwarzschild BH at a critical value of the GB coupling $\alpha_\text{cr}$, independent of the Ricci coupling $\beta$.
For sufficiently large $\beta$ the BH branches start toward smaller masses $\frac{M}{2\sqrt{\alpha}}$ (for fixed $\alpha$), necessary for radial stability \cite{Antoniou:2021zoy,Antoniou:2022agj}. 

\begin{figure}[t!]
{\hspace*{-0.5cm}\includegraphics[width=0.5\textwidth]{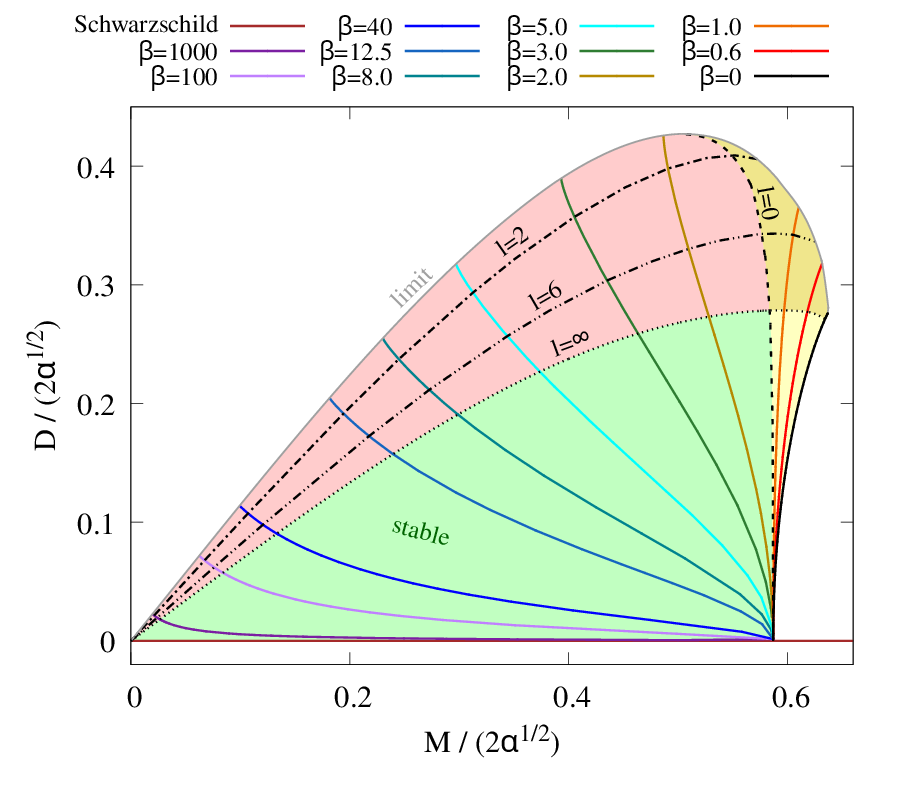}}
\vspace{-15pt}
    \caption{Branches of scalarized black holes for several values of the Ricci coupling $\beta$ {(solid colored lines)}. 
The dashed curves indicate the onset points of the spherical, quadrupole, $l=6$ and eikonal deformation instabilities.
    } 
        \label{fig1}
\end{figure}

In Fig.~\ref{fig1} we show the scalar charge against the mass of scalarized BHs, scaling by $\alpha$. 
The colored curves represent branches of scalarized BHs with fixed values of $\beta$, and the domain of existence is represented by the colored areas.
The boundaries of the domain are given by the Schwarzschild BH (horizontal brown curve), the EsGB branch ($\beta=0$, black solid curve), and the limit solutions for which the horizon constraint is saturated (grey curve).

At a branch turning point, a radial zero mode appears (which happens for values of $\beta$ below $1.87$ approximately).
When continuing the branches to larger masses {$\frac{M}{2\sqrt{\alpha}}$}, radial stability is lost.
In Fig.~\ref{fig1}, solutions possessing a radial zero mode are marked with the $l=0$ dashed black curve.
Scalarized BHs to the right of this curve are unstable (yellow and golden areas).

Previous studies have shown that radial stability does not guarantee full stability. 
Radially stable EsGBR black holes may develop a quadrupolar deformation instability \cite{Kleihaus:2023zzs}, and in the eikonal limit an angular Laplacian instability has been identified \cite{Minamitsuji:2024twp}. 
These observations suggest that the stability properties of scalarized black holes cannot be captured by a single perturbative sector. 
Here we demonstrate that these apparently disconnected instabilities are part of a broader hierarchy of nonspherical deformation instabilities extending from the eikonal regime toward progressively lower multipoles.

\prlsec{Polar perturbations.}
To this end, we now consider the polar perturbations of the metric given by
\begin{eqnarray}
 &   \delta g_{\mu\nu}(t,r,\theta,{\varphi}) = 
    Y_{lm}(\theta,{\varphi})e^{-i\omega t} \times
    \nonumber \\
 &    \begin{pmatrix}
e^{2\nu}H_0(r) & H_1(r) & 0 & 0\\
H_1(r) & e^{2\lambda} H_2(r) & 0 & 0\\
0 & 0 & r^2K(r) & 0\\
0 & 0 & 0 & r^2\sin^2\theta K(r)
\end{pmatrix} \,
,
\end{eqnarray}
and the scalar field perturbations
\begin{equation}
    \delta{\phi}(t,r,\theta,{\varphi})=u(r)Y_{lm}(\theta,{\varphi})e^{-i\omega t} \,.
\end{equation}
Because of spherical symmetry and without loss of generality in the following we fix $m=0$.

From the equations of motion follows a system of Ordinary Differential Equations (ODE's) for the perturbations that can be written in matrix form as
\begin{equation}\label{ODEs}
    \mathcal{D}_i\textbf{X}=0 \, , \;\;\;\;i=1,...,4
\end{equation}
where $\textbf{X}=(H_0,H_1,K,u)$ and $\mathcal{D}_i$ are the resulting {linear} differential operators.

We are interested in solutions describing purely outgoing waves at infinity and purely ingoing waves at the horizon. 
Hence we reparametrize the perturbation functions so that at infinity they behave like
$\textbf{X} \sim r e^{i\omega r^*} \hat{\textbf{X}}$,
and at the horizon like
$\textbf{X}\sim (r-r_H)^{-1} e^{-i\omega r^*} \hat{\textbf{X}}$,
where $\hat{\textbf{X}}$ are regular functions at the boundaries.
$r^*$ denotes the tortoise coordinate, that behaves like $\frac{dr^*}{dr} \approx 1+2M/r$ at infinity 
and 
like $\frac{dr^*}{dr} \approx g_1/(r-r_H)$ at the horizon ($g_1$ being a constant). 

The mode spectrum is obtained using a spectral decomposition of the perturbations, described earlier \cite{Blazquez-Salcedo:2023hwg,Blazquez-Salcedo:2025dit}. 
Thus, we employ a compactified radial coordinate $x=1-\frac{r_H}{r}$ and decompose the {perturbation functions} into Chebyshev polynomials
\begin{equation}\label{spectral_decomp}
    \hat{\textbf{X}}(x)=\sum_{k=0}^{N_p-1}\textbf{C}_kT_k(x) \,,
\end{equation}
where $N_p$ is the size of the grid, $\mathbf{C}_k$ are the coefficients of the decomposition and $T_k(x)$ denotes the Chebyshev polynomials of the first kind. 
The constants $C_{i,k}$ are obtained by solving the ODEs. 
We discretize $x$ into a grid based on a Gauss-Lobatto distribution. 
After substituting equation \eqref{spectral_decomp} into the corresponding equations we are left with $4\times N_p$ algebraic equations for the polar perturbations. 
This problem is written as a quadratic eigenvalue problem $\left(\mathcal{M}_0+\mathcal{M}_1\omega+\mathcal{M}_2\omega^2\right)\mathbf{C}=0$,
the $\mathcal{M}_k$ being  square matrices of size $4\times N_p$,
that is solved  using Matlab with the Advanpix Multiprecision Computing Toolbox \cite{Advanpix}. 

This scheme allows us to calculate the full quasinormal mode spectrum of the scalarized black holes, and in particular, the unstable modes for arbitrary values of $l$.

\prlsec{Angular instabilities.}

The onset  
of the 
nonspherical instabilities are indicated in Fig.~\ref{fig1} 
with the $l=\infty$ dotted black curve.
Solutions above this curve possess instabilities
for multipoles in the range $2\le l < \infty$. 
Several features are immediately apparent. 
First, no instability is present on the initial part of the radially stable branches (green colored area). 
Second, the instability appears first at large multipole number, while lower multipoles remain stable. 
Third, progressively lower multipoles become unstable only deeper along the branch.
These observations already suggest the existence of an ordered structure in multipole space.
To simplify the illustration of the hierarchy of instabilities, 
in Fig.~\ref{fig1} we also show with dot-dash curves the $l=2$ and $l=6$ zero mode curves.
Above the $l=6$ curve
the scalarized BHs possess $l\ge 6$ instabilities.
Similarly, above the $l=2$ curve the scalarized BHs possess the quadrupolar instability, and in fact solutions in this region possess all the $l\ge2$ instabilities.

Note that we do not find unstable $l=1$ modes.
However, as commented before,
the $l=0$ unstable modes can be found 
in the colored regions to the right of the $l=0$ zero mode curve (gold and light yellow).
Hence, BHs in the pink region 
and below the
$l= l_0$ zero mode curve
possess angular instabilities with 
$l>l_0$ 
($l_0\ge 2$)
while BHs in the golden region possess, in addition, the radial instability.

\begin{figure}[tb]
{\hspace*{-0.2cm}\includegraphics[width=0.75\columnwidth,angle=-90]{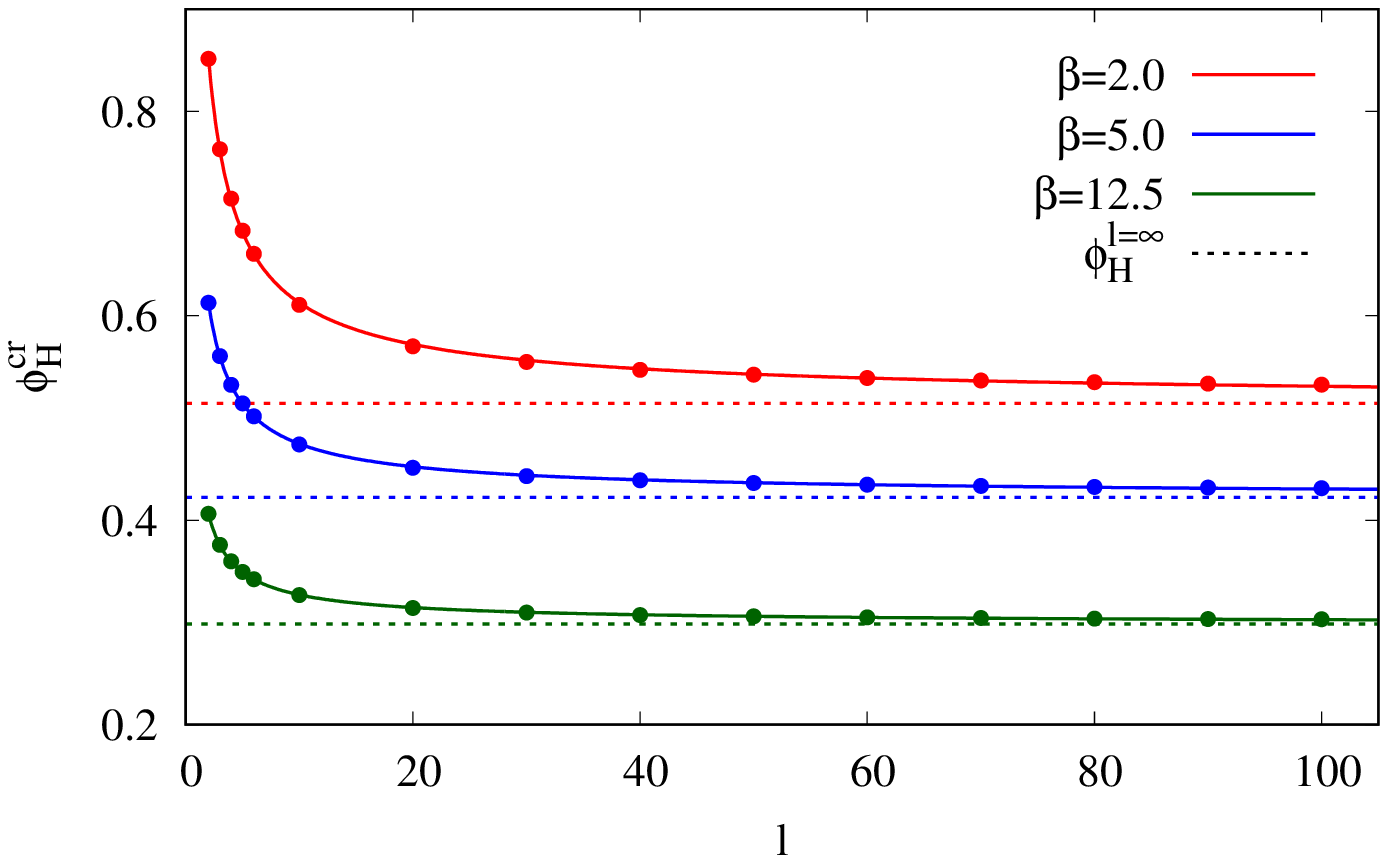}}
\vspace{-15pt}
    \caption{
Angular-instability phase diagram. 
The dots are the onset values $\phi_H^\text{cr}(l)$ for different values of $\beta$, and the curves are the corresponding fits to equation (\ref{fit_eq}).
Configurations with $\phi_H>\phi_H^\text{cr}(l)$  have an unstable mode for that multipole.
The curves approach finite large-$l$ limits ($\phi_H^{l=\infty}$) that coincide with the independently derived angular-Laplacian instability threshold \cite{Minamitsuji:2024twp}.
}
        \label{fig2}
\end{figure}

To investigate the ordered structure in multipole space,
we determine the critical horizon value  $\phi_H^\text{cr}(l)$ at which the zero mode appears for each multipole. 
The resulting angular-instability phase diagram is shown in Fig.~\ref{fig2} for $\beta=2$, $\beta=5$ and $\beta=12.5$.
For a given multipole, configurations with  $\phi_H>\phi_H^\text{cr}(l)$ are unstable.
Remarkably, the onset points 
are not merely
a collection of disconnected thresholds.
Instead, they lie on smooth curves approaching finite large-$l$ limits. 
The instability thresholds are accurately described by
\begin{equation}
\phi_H^\text{cr}(l)=\phi_H^{l=\infty}+A\,l^{-p}\,.
\label{fit_eq}
\end{equation}
The fitted parameters are summarized in Table \ref{tab1}. 
\begin{table}[h]
\caption{\label{tab1}
Fit parameters for the instability-threshold hierarchy
$\phi_H^{\rm cr}(l)=\phi_H^{l=\infty}+A\,l^{-p}$. 
}
\begin{ruledtabular}
\begin{tabular}{ccccc}
$\beta$ & $\phi_H^{l=\infty}$ & $A$ & $p$ & RMSE \\
\hline
2 & 0.51417 & 0.57630 & 0.76756 & $1.27\times10^{-3}$ \\
5 & 0.42229 & 0.33336 & 0.80343 & $5.36\times10^{-4}$ \\
12.5 & 0.29840 & 0.19148 & 0.82533 & $2.71\times10^{-4}$ \\
\end{tabular}
\end{ruledtabular}
\end{table}
The small residual errors, which decrease with increasing $\beta$, together with the comparable exponents obtained for $\beta=2$, $5$, and $12.5$, indicate that the instability thresholds obey a common scaling law over a wide range of couplings. 
The hierarchy therefore appears to be an intrinsic structural property of the scalarized branches rather than an accidental feature of individual solutions.

The finite values of $\phi_H^{l=\infty}$ define the boundary of the angularly stable region of the branch. 
The hierarchy therefore provides a quantitative phase diagram for the onset of nonspherical instabilities.
For $\phi_H<\phi_H^{l=\infty}$, the branch is stable against all nonspherical modes investigated here, whereas for $\phi_H>\phi_H^{l=\infty}$ instability first appears in the large-$l$ sector and subsequently extends toward lower multipoles.
The hierarchy therefore organizes the onset of nonspherical instabilities across the angular sectors of the theory.

No unstable dipole mode was found in the parameter range investigated. 
The hierarchy therefore starts at the quadrupolar sector and extends continuously toward the eikonal regime. 
This indicates that the dipole sector is dynamically distinct from the family of deformation instabilities studied here.

The asymptotic values extracted from the hierarchy coincide with the independently derived angular-Laplacian instability threshold \cite{Minamitsuji:2024twp}. 
The finite threshold curves of Fig.~\ref{fig2} therefore provide the missing connection between the previously identified quadrupole instability \cite{Kleihaus:2023zzs} and the asymptotic angular-Laplacian instability \cite{Minamitsuji:2024twp}, showing that they arise as different manifestations of a continuous hierarchy of instability thresholds.
This strongly suggests that the eikonal instability represents the limiting member of the hierarchy rather than an unrelated phenomenon. 

\begin{figure}[b]
{\hspace*{-0.2cm}
\includegraphics[width=0.75\columnwidth,angle=-90]
{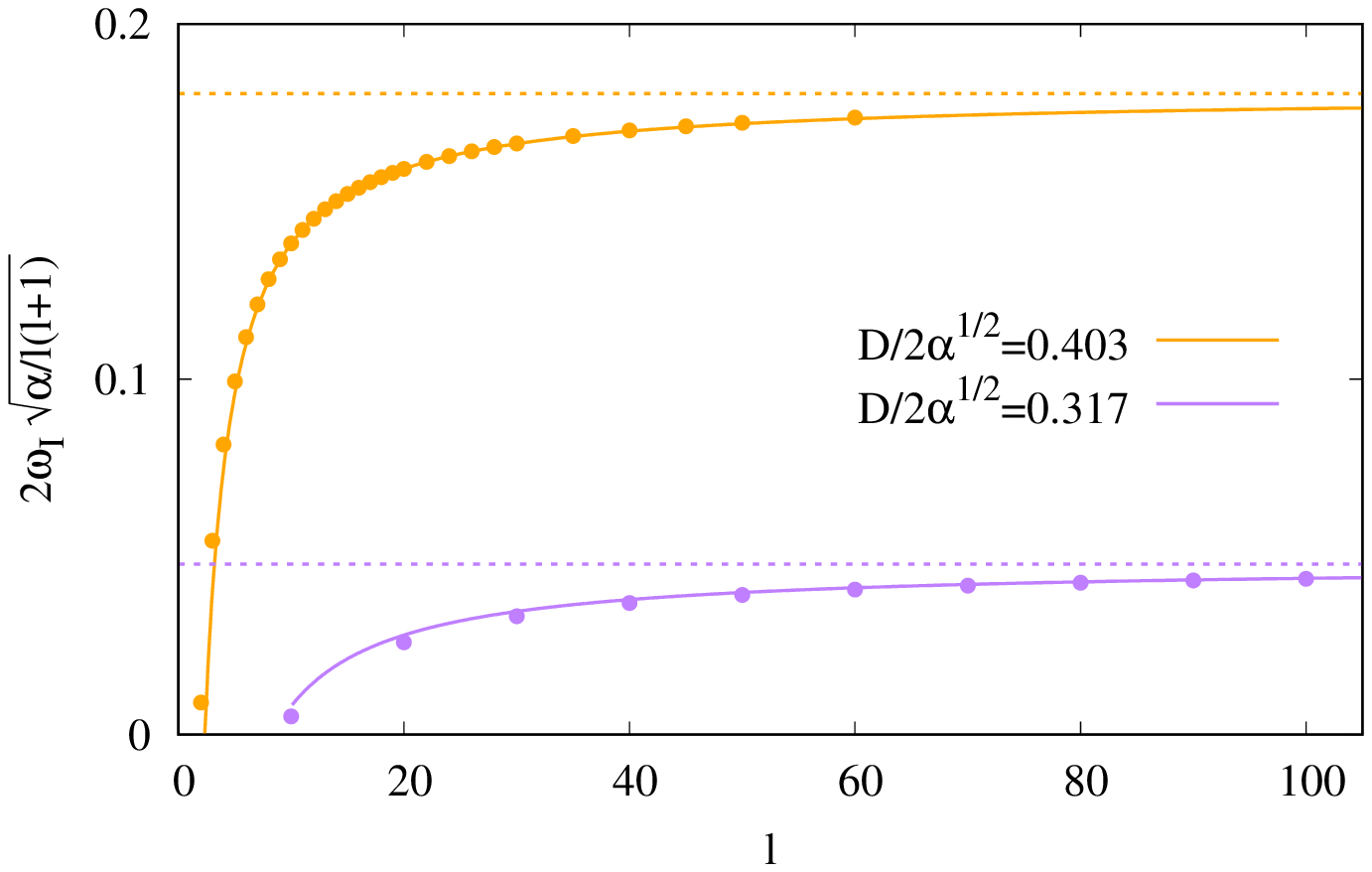}}
\vspace{-15pt}
\caption{
Eikonal scaling of the unstable modes for two solutions on the $\beta=2$ branch,
scaling with the coupling constant $\alpha$.
The solution with larger scalar charge already possesses an unstable quadrupole mode, whereas the second solution exhibits only higher-multipole instabilities. 
In both cases $\omega_I/\sqrt{l(l+1)}$ approaches a constant value for large $l$, demonstrating the persistence of eikonal scaling across different stages of the instability hierarchy.}
\label{fig3}
\end{figure}

Independent support for this interpretation is provided by the unstable mode frequencies. 
Figure \ref{fig3} shows that the growth rate satisfies
\begin{equation}
\frac{\omega_I(l)}{\sqrt{l(l+1)}}\rightarrow {\rm const.}
\end{equation}
for large multipole number. 
To illustrate this behavior, we consider two solutions on the same $\beta=2$ branch. 
One solution lies in the region where only higher multipoles are unstable, whereas the second lies deeper along the branch and already possesses an unstable quadrupole mode.

The large-$l$ data are well described by an inverse-$l$ approach to a constant,
\begin{equation}
\frac{\omega_I(l)}{\sqrt{l(l+1)}} = a+\frac{b}{l}\,,
\end{equation}
which we use only as an empirical parametrization of the asymptotic limit.
The same asymptotic behavior is obtained both before and after the appearance of the quadrupole instability, demonstrating that the eikonal scaling persists across
different stages of the hierarchy.
This form accurately fits both solutions shown in Fig.~\ref{fig3}, with small residual errors, although the asymptotic constant $a$ depends on the position along the branch (i.e. depends on the scalar charge). 
Thus the unstable modes exhibit the characteristic eikonal scaling expected from the angular-Laplacian instability.
Together, the finite limiting values of the onset thresholds in Fig.~\ref{fig2} and the frequency scaling in Fig.~\ref{fig3} identify the large-$l$ end of the hierarchy with an eikonal instability.

The same hierarchy can be visualized directly on the scalarized branches.
Figure~4 shows representative branches in the $(\phi_H,A_H)$ plane, together with the corresponding zero-mode points. 
Along each branch, the large-$l$ instability is encountered first, while the onset of lower multipoles occurs only deeper along the branch. 
This provides a geometric representation of the progression from the eikonal regime
toward the quadrupole sector.

\begin{figure}[tb]
{\hspace*{-0.2cm}\includegraphics[width=0.75\columnwidth,angle=-90]{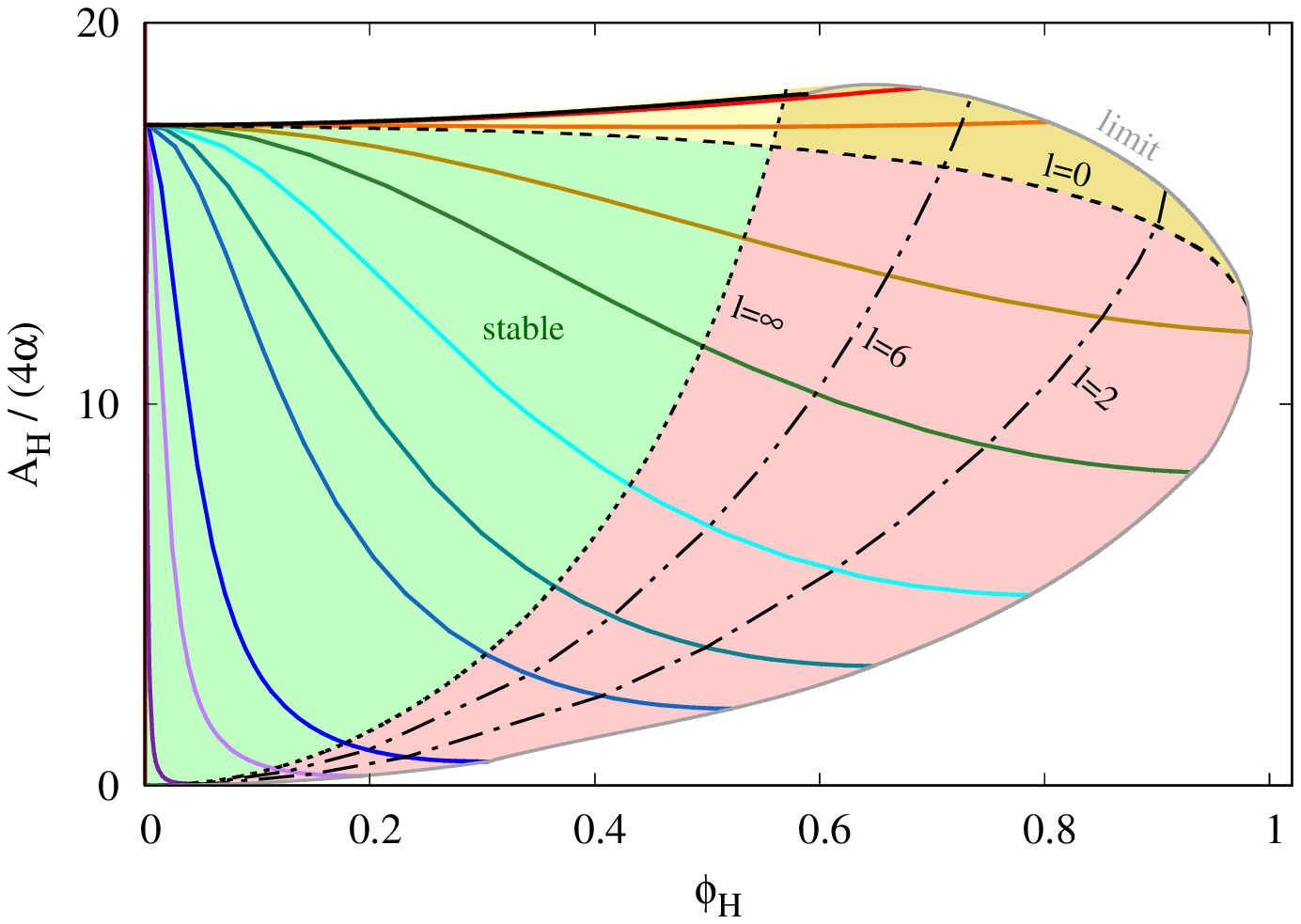}}
\vspace{-15pt}
    \caption{Scalarized black-hole branches in the $(\phi_H,A_H)$ plane.
    Markers indicate the onset of the radial and nonspherical zero modes.
The ordering of the markers along the branches illustrates the hierarchy of angular instabilities: the large-$l$ instability appears first, followed by progressively lower multipoles.
}
        \label{fig4}
\end{figure}

\prlsec{Conclusions.}
We have investigated the stability of scalarized black holes in Einstein-scalar-Gauss-Bonnet-Ricci theory and uncovered a previously unexplored organization of nonspherical instabilities along their fundamental branches.

Starting from an angularly stable region, instability first appears in the eikonal regime and subsequently extends toward progressively lower multipoles, forming an ordered hierarchy of deformation instabilities. 
The corresponding onset points lie on smooth threshold curves that obey a common scaling law and approach finite large-$l$ limits, while the dipole sector remains stable throughout the domain investigated. 
The resulting phase diagram defines the boundary of the angularly stable region and connects the previously known quadrupole and angular-Laplacian instabilities through a continuous hierarchy of instability thresholds. 
In contrast, radial stability changes only at the turning points of the branches.

More generally, our findings raise the possibility that similar hierarchical structures may arise in other gravitational systems with continuous families of solutions. 
A particularly interesting direction is the extension to rotation. 
Rotating black holes in Einstein-scalar-Gauss-Bonnet theory are known to exhibit
a rich phase structure, including spinning scalarized and excited solutions, as well as spin-induced scalarization 
\cite{Cunha:2019dwb,Collodel:2019kkx,Dima:2020yac,Hod:2020jjy,Doneva:2020nbb,Herdeiro:2020wei,Berti:2020kgk}.
It would therefore be important to determine how the hierarchy of angular instabilities identified here generalizes to rotating scalarized black holes and whether it provides an organizing principle for the various rotating families that bifurcate from scalarized black-hole branches.

\prlsec{Acknowledgments.}
We would like to thank Burkhard Kleihaus for his comments on the manuscript.
We gratefully acknowledge support by MICINN project PID2021-125617NB-I00 ``QuasiMode''.
JLBS gratefully acknowledges support from MICINN project CNS2023-144089 ``Quasinormal modes''.
FSK gratefully acknowledges support from ``Atracci\'on de Talento Investigador Cesar Nombela'' of the Comunidad de Madrid under the grant number 2024-T1/COM-31385.
PNM gratefully acknowledges support from Universidad Complutense de Madrid through ``Contratos predoctorales
de personal investigador en formación CT25/24''. 
PNM thanks the University of Oldenburg for hospitality during part of this project.

\end{document}